# Spherical Symmetric Solution in f(R) Model Around Charged Black Hole

$^a$**A. Aghamohammadi**[1], $^{b,c}$**Kh. Saaidi**[2], $^a$**M. R. Abolhasani**[3],
$^b$**A. Vajdi**[4]

$^a$*Plasma Physics Research Center, Science and Research Branch Islamic Azad University of Tehran, Iran*
$^b$*Department of Physics, Faculty of Science, University of Kurdistan, Pasdaran Ave., Sanandaj, Iran*
*Faculty of Science, Azad University of Sanandaj, Sanandaj, Iran*

**Abstract**

A static, asymptotically flat, spherically symmetric solutions is investigated in $f(R)$ theories of gravity for a charged black hole. We have studied the weak field limit of $f(R)$ gravity for the some $f(R)$ model such as $f(R) = R + \epsilon h(R)$. In particular, we consider the case $\lim_{R \to o} h(R)/h'(R) \to 0$ and find the space time metric for $f(R) = R + \frac{\mu^4}{R}$ and $f(R) = R^{1+\epsilon}$ theories of gravity far away a charged mass point.

---

[1]agha35484@yahoo.com
[2]ksaaidi@uok.ac.ir
[3]mrhasani@modares.ac.ir
[4]Avajdi@uok.ac.ir

# 1 Introductions

The acceleration expansion of the Universe is one of open problems in the cosmology [1]. This phenomena is not explained by general relativity theory. Some of people think this acceleration expansion my be due to some unknown energy- momentum component which having the equation of state as $p = \omega\rho$. This opinions led to several theoretical models such as quintessential scenarios [2] which generalize the cosmological constant approach [3], higher dimensional scenarios [4, 5] or the resort to cosmological fluids with exotic equation of state [6]. Combining all of the observations and above opinions, a cosmological model has emerged, a universe dominated by cosmological constant- like dark energy and cold dark matter. One obvious contender for the role of dark energy is Einstein's cosmological constant, but particle physics failed to predict the correct density. Whereas **GR** theory con not explains the accelerate equation of universe, over the last decades, a wide number of approaches have been developed to generalized it. The simplest way of generalizing **GR** theory of gravities to consider a gravitational action described by a function of Ricci scalar $f(R)$ instead of the Einstein - Hilbert action. Generalized $f(R)$ gravity may be considered a reliable mechanism to explain the acceleration equation of universe [6-23]. One of the initiative $f(R)$ model supposed to explaining the positive acceleration of expanding universe has $f(R)$ action as $f(R) = R - \mu^4/R$ [24]. In this model, for large values of Ricci scalar, $R \gg \mu^2$, $f(R)$ function tends to $f(R) = R$, so we expect for these values of $R$ the modification become negligible but for small values of Ricci scalar, $R \lesssim \mu^2$, $f(R)$ action is not the linear one thus for this values of Ricci scalar gravity is modified. Because for large values of $R$ the modification is negligible, this model can not explain the inflation but there is several viable models unifying inflation and late time acceleration[25].

After proposing the $f(R) = R - \mu^4/R$ model it was appeared this model suffer several problems. In the metric formalism, initially Dolgov an Kawasaki discovered the violent instability in the matter sector [26]. The analysis of this instability generalized to arbitrary $f(R)$ models [27, 28] and it was shown an $f(R)$ model is stable if $d^2f/dR^2 > 0$ and unstable if $d^2f/dR^2 < 0$. Thus we can deduce $R - \mu^4/R$ suffer the Dolgov-Kawasaki instability but this instability removes in the $R + \mu^4/R$ model, where $\mu^4 > 0$. Furthermore one can see in the $R - \mu^4/R$ model the cosmology is inconsistent with observation when non-relativistic matter is present. In fact there is no matter dominant era [29, 30]. However the recent study shows the standard epoch of matter domination can be obtained in the $R + \mu^4/R$ model [31].

Spherically symmetric stars in the context of $f(R)$ models of gravity have been studied in [32-37]. In the Platini formalism the spherically symmetric solution is Schwarzschild -de'Sitter metric with effective cosmological constant and in the metric formalism the spherically symmetric solution of $f(R)$ theory of gravity suffers from a low - mass equivalent scalar field, that is incomparable with solar system test of general relativity [38, 39].

The present paper analysis is based on the weak field limit of spherically symmetric solution of $f(R)$ gravity model at the metric approach. Up to now, the discussion on the weak field limit of $f(R)$ theory of gravity is done and there are few papers which claiming different results [27, 32, 33, 34, 40, 41, 42, 43]. In this work we obtain the generalized Reissner- Nordestrom solution for a charged mass point state in a general $f(R)$ gravity model. Indeed the Reissner-Nordstrom solution for a charged mass point is in many ways similar to that of the more complicated Kerr solution describing rotating black hole, as well as the metric fluctuation to the charge black holes cause the Hawking effect, but the approach that we have adapted is that look for a static asymptotically flat in the $f(R)$ theory and investigated if the $f(R)$ theory solution end up to solution of the Einstein-Maxwell field equation.

The scheme of the paper is the following. In Sec.2 we review field equation of $f(R)$ gravity in the metric approach. We study the static, asymptotically flat, spherically symmetric solutions of $f(R)$ gravity far away a charged black hole and obtain the generalized Einstein-Maxwell equation. In Sec.3 we discuss some viable $f(R)$ models in the present our work and we obtain the generalized solution far from of a charged mass point.

## 2 Theoretical Framework

In this section, we offer to investigate the Reissner- Nordestrom solution for a charged mass point in the $f(R)$ gravity model. We consider an action as

$$S = \int \sqrt{-g} \left[ \frac{f(R)}{2} + k(L_m + L_{rad}) \right] d^4x. \tag{1}$$

Here R is a Ricci scalar, $L_m$ is matter Lagrangian and $L_{rad}$ is electromagnetic field (EM) Lagrangian. We consider a class of modified gravity such as $f(R) = R + \epsilon h(R)$, which is obtain by adding a perturbative function $\epsilon h(R)$, to the Einstein-Hilbert action, where $\epsilon$ is a small parameter. In this case the field equation, using the metric approach, can be derived from (1) as,

$$G_{\mu\nu} = -\epsilon \left[ G_{\mu\nu} + g_{\mu\nu}\Box - \nabla_\mu \nabla_\nu + \frac{g_{\mu\nu}}{2}\left(R - \frac{h(R)}{\varphi(R)}\right)\right]\varphi(R) + k\tilde{T}_{\mu\nu}, \tag{2}$$

where $G_{\mu\nu}$ and $\tilde{T}_{\mu\nu} = T^m_{\mu\nu} + T^{em}_{\mu\nu}$ are Einstein and stress-energy tensors respectively, $\Box \equiv \nabla_\alpha \nabla^\alpha$ and $\varphi(R) = dh(R)/dR$. We look for a static, asymptotically flat, spherically symmetric solution of the modified Einstein-Maxwell field equations (2). By contracting field equation (2) we have

$$R = \epsilon \left[R - \frac{2h(R)}{\varphi(R)} + 3\Box\right]\varphi(R) - k\tilde{T}. \tag{3}$$

where $\tilde{T}$ is the trace of $\tilde{T}_{\mu\nu}$. It is clearly seen that for $\epsilon \to 0$ the equation (2) and (3) reduces to the equivalent equation in Einstein general relativity theory. Furthermore it is seen that as $\epsilon \to 0$, corresponding to the general relativity, $G_{\mu\nu}$ and $R$ tend to zero. Hence, we can neglected $\epsilon G\mu\nu$ and $\epsilon R$ in (2) and (3). In addition, the Maxwell tensor $F_{\mu\nu}$, must satisfy Maxwell's equations in out of sphere, i.e.;

$$\nabla_\nu F^{\mu\nu} = 0, \tag{4}$$

$$\partial_{[\mu} F_{\nu\gamma]} = 0. \tag{5}$$

The assumptions of static, asymptotically flat, spherically symmetric solution of the modified Einstein- Maxwell field equations, means that one can introduce the line element as

$$ds^2 = -(1 + a(r))dt^2 + \frac{1}{(1 + b(r))}dr^2 + r^2 d\Omega^2. \tag{6}$$

Hence, the components of Ricci tensor corresponding with the line element, can be written in terms of $a(r)$ and $b(r)$, as

$$R^t_t = \frac{a'}{r} + \frac{a''}{2}, \tag{7}$$

$$R^r_r = \frac{b'}{r} + \frac{a''}{2}. \tag{8}$$

$$R^\theta_\theta = R^\phi_\phi = \frac{b}{r^2} + \frac{a'}{2r} + \frac{b'}{2r}, \tag{9}$$

and one can obtain the Ricci scalar as

$$R = a'' + 2(\frac{a'}{r} + \frac{b'}{r} + \frac{b}{r^2}). \tag{10}$$

The components of the Einstein tensor in terms of parameters of the metric, $a(r)$ and $b(r)$, can be find as following

$$G^t_t = -\frac{b'}{r} - \frac{b}{r^2}, \tag{11}$$

$$G^r_r = -\frac{a'}{r} - \frac{b}{r^2}, \tag{12}$$

$$G^\theta_\theta = G^\phi_\phi = -\frac{a''}{2} - \frac{a'}{2r} - \frac{b'}{2r}. \tag{13}$$

On the other hand, by using equations (2) and (6), we can obtain the components of Einstein tensor. But in the first, we must obtain the tensor of energy momentum. In the out of matter we have only the Maxwell's tensor, which is as

$$T^\mu_\nu = \frac{1}{4\pi}\left(\frac{1}{4}\delta^\mu_\nu F_{cd}F^{cd} - g^{cd}F^\mu_c F_{\nu d}\right). \tag{14}$$

We assume that the charge distribution has spherical symmetry. This case is equivalent with the state which we take a point charge particle to be situated at the origin of the coordinates. Moreover, the charged particle will give rise to an electrostatic field which is purely radial. This means that the Maxwell tensor must take on the form

$$F_{\mu\nu} = E(r)\begin{pmatrix} 0 & -1 & 0 & 0 \\ 1 & 0 & 0 & 0 \\ 0 & 0 & 0 & 0 \\ 0 & 0 & 0 & 0 \end{pmatrix} \tag{15}$$

By using equations (6), (14) and (15) we find the components of energy momentum tensor as

$$kT^t_t = G(1+b-a)\frac{q^2}{r^4}, \tag{16}$$

$$kT^r_r = G(1+b-a)\frac{q^2}{r^4}, \tag{17}$$

$$kT^\theta_\theta = kT^\phi_\phi = -G(1+b-a)\frac{q^2}{r^4}, \tag{18}$$

where we use $E = \frac{q}{r^2}$ and set $k = 8\pi G$ ( $G$ is the gravitational constant). Using equations (6), (36), (17)and (18) we can obtain the Einstein tensor as

$$G^\nu_\mu = -\epsilon \left(\delta^\nu_\mu \Box - \nabla^\nu \nabla_\mu\right) \varphi(R) + kT^\nu_\mu, \tag{19}$$

$$G^t_t = G(1-a+b)\frac{q^2}{r^4} - \epsilon\left(\varphi'' + 2\frac{\varphi'}{r}\right), \tag{20}$$

$$G^r_r = G(1-a+b)\frac{q^2}{r^4} - 2\epsilon\frac{\varphi'}{r}, \tag{21}$$

$$G^\theta_\theta = G^\phi_\phi = -G(1+b-a)\frac{q^2}{r^4} - \epsilon\left(\varphi'' + \frac{\varphi'}{r}\right). \tag{22}$$

where $\varphi = \frac{dh}{dR}$ and $(')$ denotes the differentiation with respect to $r$. Therefore, using the two set of equations (11, 12, 13) and (20, 21, 22), we can arrive at

$$-\frac{b'}{r} - \frac{b}{r^2} = G\frac{q^2}{r^4} - \epsilon\left(\varphi'' + 2\frac{\varphi'}{r}\right), \tag{23}$$

$$-\frac{a'}{r} - \frac{b}{r^2} = G\frac{q^2}{2r^4} - 2\epsilon\frac{\varphi'}{r}, \tag{24}$$

$$-\frac{a''}{2} - \frac{a'}{2r} - \frac{b'}{2r} = -G\frac{q^2}{r^4} - \epsilon\left(\varphi'' + \frac{\varphi'}{r}\right). \tag{25}$$

For obtaining equations (23, 24, 25), we neglect terms $aq^2$, $bq^2$, $a^2$, $b^2$.

## 3  $f(R)$ Models

In this section we shall completed the our analyzes for some models of $f(R)$ theory in the outside a charged star. It is well known that the energy momentum tensor in outside a charged star is as (14). It is clearly seen that its trace is zero. From Eqs. (2, 3) it is clear that as $\epsilon \to 0$, corresponding to the general relativity, $G_{\mu\nu}$ and $R$ tend to zero. Hence, in r.h.s of Eqs.(2, 3) we may neglect $\epsilon G_{\mu\nu}$ and $\epsilon R$. Furthermore if $h(R)/\varphi(R)$ vanishes as $R \to 0$, i.e.;

$$\lim_{R \to 0} \frac{h(R)}{\varphi(R)} \to 0, \tag{26}$$

then in Eqs.(2, 3) we may neglect the $h(R)/\varphi(R)$ term. For many $f(R)$ models, for example $h(R) = 1/R$, $R \ln R$, $\ln R$, this condition is satisfied. So, assuming the condition (6), we may rewrite trace equation (3) as

$$R = 3\epsilon \nabla^2 \varphi(R), \tag{27}$$

### 3.1  $f(R) = R + \mu^4/R$ Model

For this model we may write

$$\epsilon h(R) = \frac{\mu^4}{R}, \tag{28}$$

$$\epsilon \varphi(R) = -\frac{\mu^4}{R^2}, \tag{29}$$

which $h(R)$ satisfy the condition (26). Inserting $\varphi(R)$ from the above equation in to the trace equation (27) we obtain

$$R = 7\alpha \mu^{\frac{4}{3}} r^{-\frac{2}{3}}, \tag{30}$$

$$\epsilon \varphi = -\frac{1}{49\alpha^2} \mu^{\frac{4}{3}} r^{\frac{4}{3}}, \tag{31}$$

where $\alpha^3 = 4/147$. Inserting $\epsilon \varphi$ from the above equation in to Eqs.(23), (24) and (25) and introducing $a$ and $b$ as

$$a = \sum_n A_n r^n + \sum_m \frac{B_m}{r^m}, \tag{32}$$

$$b = \sum_l C_l r^l + \sum_s \frac{D_s}{r^s} \tag{33}$$

we can obtain

$$a = -\frac{2M}{r} + \frac{Gq^2}{r^2} - \frac{3}{4} \alpha \mu^{\frac{4}{3}} r^{\frac{4}{3}}, \tag{34}$$

$$b = -\frac{2M}{r} + \frac{Gq^2}{r^2} - \alpha \mu^{\frac{4}{3}} r^{\frac{4}{3}}. \tag{35}$$

where $M$ is a constant which is obtained from matching the interior and exterior solutions of field equations. Therefore the space time metric for outside of a charged mass point in this model is

$$ds^2 = -(1 - \frac{2M}{r} + \frac{Gq^2}{r^2} - \frac{3}{4}\alpha\mu^{\frac{4}{3}}r^{\frac{4}{3}})dt^2 \tag{36}$$
$$+ (1 - \frac{2M}{r} + \frac{Gq^2}{r^2} - \alpha\mu^{\frac{4}{3}}r^{\frac{4}{3}})^{-1}dr^2 + r^2d\Omega^2,$$

### 3.2 $f(R) = R^{1+\epsilon}$ Model

Since we are interested in the limit $\epsilon \to 0$ we can expand $f(R) = R^{1+\epsilon}$ around $\epsilon = 0$. Hence $f(R) = R + \epsilon R \ln R$, and definitions lead to

$$h(R) = R \ln R, \tag{37}$$
$$\varphi(R) = 1 + \ln R. \tag{38}$$

It is clear that $h(R)$ satisfy the condition (26). Inserting $\varphi(R)$ from Eq.(38) in the trace equation (3), we obtain an equation for Ricci scalar. Solving the obtained equation leads to

$$R = \frac{6\epsilon}{r^2}. \tag{39}$$

Inserting $R$ from the above equation in (38) and then in the r.h.s of Eqs (23), (24), (25) and use the expressions (32), (33) gives $a, b$ as

$$a = -\frac{2M}{r} + \frac{Gq^2}{r^2} + 2\epsilon \ln r, \qquad b = -\frac{2M}{r} + \frac{Gq^2}{r^2} + 2\epsilon, \tag{40}$$

where $M$ is a constant which is obtained from matching the external and internal solution of field equation. Therefore the space time metric for outside of the charged mass point space in this model is

$$ds^2 = -(1 - \frac{2M}{r} + \frac{Gq^2}{r^2} + 2\epsilon \ln r)dt^2 \tag{41}$$
$$+ (1 - \frac{2M}{r} + \frac{Gq^2}{r^2} + 2\epsilon)^{-1}dr^2 + r^2d\Omega^2,$$

It is obviously seen that when $\mu$ or $\epsilon$ tends to zero equations (36) and (41) reduces to ordinary Reissner- Nordestrom metric

$$ds^2 = -(1 - \frac{2M}{r} + \frac{Gq^2}{r^2})dt^2 \tag{42}$$
$$+ (1 - \frac{2M}{r} + \frac{Gq^2}{r^2})^{-1}dr^2 + r^2d\Omega^2,$$

This metric has two possible horizon which can be found as follows,

$$r = M \pm \sqrt{M^2 - G^2 q^2}, \qquad (M^2 \geq G^2 q^2). \qquad (43)$$

These two values are due to charge $q$. In fact, when a black hole becomes charged, the event horizon shrinks, and another one appears, near the singularity [44].

## 4 Discussion and summary

In this paper, we have studied the spherically symmetric solution of some viable $f(R)$ models of gravity in the weak field limit at far away charged black hole and have derived that $f(R)$ gravity easily gives rise to the line element solution of Reissner- Nordestrom metric. In the metric formalism, we studied, some $f(R)$ models such as $f(R) = R + \epsilon h(R)$. We have been obtained the generalized metric for some models which have satisfied the condition $\lim_{R \to o} \frac{h(R)}{h'(R)} \to 0$. In particular, we have been obtained the generalized Reissner- Nordestrom metric for $f(R) = R + \frac{\mu^4}{R}$ and $f(R) = R^{1-\epsilon}$ theories of gravity in far away of a charged mass.

# References

[1] A. G. Riess *et al.*, Astron. J. **116**, 1009 (1998); S. Perlmutter *et al.*, Astrophys. J. **517**, 565 (1999).

[2] R. R. Caldwell, R. Dave, P. J. Steinhardt, Phys. Rev. Lett. **80**, 1582 (1998).

[3] V. Sahni, A. A. Starobinsky, Int. J. Mod. Phys. **D 9**, 373 (2000)

[4] R. Maartens, Living Rev. Rel. **7**, 7 (2004).

[5] M. Pietroni, Phys. Rev. **D 67**, 103523 (2003).

[6] M. C. Bento, O. Bertolami, A. A. Sen, Phys. Rev. **D 67**, 063003 (2003).

[7] S. Capozziello, *et al.*, Int. J. Mod. Phys. **D 11**, 483 (2002).

[8] S. Nojiri, S. D. Odintsov, Phys. Lett. **B 576**, 5 (2003).

[9] S. Nojiri and S. D. Odintsov, Phys. Rev **D 68**, 123512 (2003).

[10] D. N. Vollick, Phys. Rev. **D 68**, 063510 (2003)

[11] S. Capozziello, S. Carloni, and A. Troisi, Phy. Rev, **D 71**, 043503 (2005).

[12] M. C. B. Abdalla, S. D. Odintsov, Class. Qunt. Grav. **22**, L35 (2005).

[13] S. M. Carroll, Phy. Rev, **D 71**, 063513 (2005).

[14] J. D. Evans and L. M. H. Hall and P. Caillo, Phy. Rev, **D 71**, 063513 (2005).

[15] G. Dvali and M. S. Turner, Astro. Phy. 0201033 (2002).

[16] S. Nojiri, S. D. Odintsov, Gen. Rel. Grav **36**, 1765 (2004).

[17] N. Arkani-Hamed, H-c. Cheng, M. A. Luty and S. Mukohyama, JHEP **05**, 043528 (2004).

[18] S. Nojiri and S. D. Odintsov, Phys. Rev **D 68**, 123512 (2003).

[19] S. A. Appleby and R. A. Battye, Phy. Lett. **B 654**, 7 (2007).

[20] L. Amendola, R. Gannouji, D. Polarski and S. Tsujikawa, Phys. Rev **D 75**, 083504 (2007).

[21] L. Amendola, D. Polarski and S. Tsujikawa, Phy.Rev. Lett. **98**, 131302 (2007).

[22] S. Capozziello, S. Carloni, and A. Troisi, Astron. Astrophys. **1**, 625 (2003).

[23] M. E. Soussa and R. P. Woodard, Gen. Rel. Grav. **36**, (2003)

[24] S. M. Carroll,*et al.*, Phy. Rev, **D 70**, 043528 (2004).

[25] S. Nojiri, S. D. Odintsov, Phys. lett. B **657**, 238 (2007); S. Nojiri, S. D. Odintsov, arxiv:0710.1738.

[26] A. D. Dolgov, M. Kawasaki, Phys. Lett. **B 573**, 1 (2003).

[27] V. Faraoni, Phys. Rev **D 74**, 023529 (2006).

[28] I. Sawicki, W. Hu, Phys. Rev. **D 75**, 127502 (2007).

[29] L. Amendola, R. Gannouji, D. Polarski, S. Tsujikawa, Phys. Rev. **D 75**, 083504 (2007); L. Amendola, D. Polarski, S. Tsujikawa, Phys. Rev. Lett**98**, 131302 (2007)

[30] S. Capozziello, S. Nojiri, S. D. Odintsov, A.Troisi, Phys. lett. B **639**, 135 (2006)

[31] J. D. Evans, Lisa. M. H. Hall, P. Caillol , Phys. Rev. **D 77**, 083514 (2008).

[32] T. Chiba, Phy. Lett. **B 575**, 1 (2003).

[33] T. Chiba, A. L. Erickcek, T. L. Smith, Phys. Rev **D 75**, 124014 (2007).

[34] T. Clifton, J. D. Barrow, Phys. Rev **D 77**, 083514 (2008).

[35] T. Muitamaki, I. Vilja, Phys. Rev **D 74**, 064022 (2006).

[36] I. Sawicki, W. Hu, Phys. Rev **D 76**, 064004 (2007).

[37] T. Faulkner, *et al.*, Phys. Rev **D 76**, 063505 (2007).

[38] G. J. Olmo, Phys. Rev. Lett. **95**, 261102 (2005).

[39] A. L. Erickcek, T. L. Smith, M. Kamionkowski, Phys. Rev. **D 74**, 121501 (2006)

[40] E. E. Flanagan, Class. Quant. Grav.**21**, 417(2003).

[41] T. P. Sotiriou, Gen. Rel. Grav. **38**, 1407 (2006).

[42] G. Allemandi, *et al.*, Gen. Rel. Grav. **37**, 1407 (2005).

[43] S. Nojiri, S. D. Odintsov, Phys. Lett. **B 657**, 238 (2007).

[44] K. Nozari, B. Fazlpour; ActaPhys. Polon. B**39**,1363 (2008).